# ARTICLE

# DNA—Au (111) Interactions and Transverse Charge Transport Properties for DNA-Based Electronic Devices

Busra Demir[a,b,c,‡], Hashem Mohammad[d,‡], M. P. Anantram[c] and Ersin Emre Oren[a,b,*]

DNA's charge transfer and self-assembly characteristics have made it a hallmark of molecular electronics for the past two decades. A fast and efficient charge transfer mechanism with programmable properties using DNA nanostructures is required for DNA-based nanoelectronics applications and devices. The ability to integrate DNA with inorganic substrates becomes critical in this process. Such integrations may effect the conformation of DNA, altering the charge transport properties. Thus, using molecular dynamics simulations and first-principles calculations in conjunction with Green's function approach, we explore the impact of Au (111) substrate on the conformation of DNA and analyze its effect on the charge transport. Our results indicate that DNA sequence, leading its' molecular conformation on Au substrate, is critical to engineer charge transport properties. We demonstrate that DNA can fluctuate on a gold substrate, sampling various distinct conformations over time. The energy levels, spatial locations of molecular orbitals and the DNA/Au contact atoms can differ between these distinct conformations. Depending on the sequence, at HOMO, the charge transmission differs up to 60 times between the top ten conformations. We demonstrate that the relative positions of the nucleobases are critical in determining the conformations and the coupling between orbitals. We anticipate these results can be extended to other inorganic surfaces and pave the way for understanding DNA inorganic interface interaction for future DNA-based electronic devices.

## Introduction

The main goal of DNA nanotechnology is to utilize DNA, a genetic substance, as a programmable structural material. DNA consists of four nucleobases: adenine (A), guanine (G), cytosine (C), and thymine (T), which are responsible for DNA's intrinsic properties such as stability, flexibility, and polymorphic structure. Their specific interactions allow DNA to be assembled into complex nanostructures that help overcome nanoscale fabrication challenges.[1] Besides, DNA nucleotides and sugar phosphates can be chemically tuned by functionalization.[2] These properties make DNA a great candidate for molecular electronics research.[3] Charge transmission along a DNA molecule has been studied for more than two decades now. Both experimental[4–10] and theoretical studies[11–15] demonstrated the feasibility of transmission along DNA molecules via overlapping pi-pi orbitals of the stacked bases and interaction with external molecules.[9,13,15–17] The charge transport in DNA molecules is reported to be driven by decoherence added to the transmission resulting from an underlying Hamiltonian by various studies.[12,13,18] In contrast, other models have proposed transport driven by rate equations in the context of tunneling and hopping between DNA bases.[4,19,20] Guanine, among the four DNA bases, is found to be the key regulator of this process, as it possesses the highest occupied molecular orbital (HOMO) level that is closest to the Fermi level of the gold electrodes.

Controllable conductivity, mechanical integrity, and structural stability are among the critical elements that molecular nanodevices should possess. DNA, like other biomolecules, is fluctuating and interacting with its surrounding environment. Several studies have been conducted to understand the relationship between conformational changes and the charge transmission along DNA molecules. Woiczikowski *et al*.[21] reported that the effect of structural fluctuations can be different based on the DNA's sequence and the base pair dynamics are important in determining the charge transport properties. Artés *et al*.[22] demonstrated that the conductance of DNA molecules increases nearly one order of magnitude when its conformation is changed from B-form to A-form. Bruot *et al*.[23] outlined that conductivity is significantly responsive to mechanical stretching and has a weak dependence on the DNA's length. Saientan Bag *et al*.[14] noted pulling direction plays a significant role in changing the conformation of DNA and therefore the conductance.

DNA is subjected to conformational changes when it is in contact with a substrate. Since DNA would be in interaction with a substrate in DNA-based nanodevices, the conformational changes due to substrate interaction would be a key element in

*Department of Materials Science & Nanotechnology Engineering, TOBB University of Economics and Technology, Ankara, Turkiye.*
*Bionanodesign Laboratory, Department of Biomedical Engineering, TOBB University of Economics and Technology, Ankara, Turkiye.*
*Department of Electrical and Computer Engineering, University of Washington, 98195 Seattle, WA, USA*
*Department of Electrical Engineering, Kuwait University, P.O. Box 5969, Safat 13060, Kuwait*
[‡] These authors contributed equally.
[*] Corresponding author.

determining the electronic properties of DNA-based nanodevices. Although there was a great effort to understand the effect of structural fluctuations on the electronic properties as mentioned above, the effect of structural changes due to an external substrate on electronic properties remains an open subject. Depending on the sequence, length, and environmental conditions, the interaction between biomolecules and inorganic substrates can change. This affects the number of interacting atoms and creates a variety of interfaces. This should in turn impact both the charge couplings between DNA and substrate atoms and between the bases. Therefore, it is important to explore the variation in conductance values depending on the conformation on the substrate.

In this study, we examine the electronic properties of DNA when in contact with an Au substrate. The Au substrate is chosen due to its chemical inertness and biocompatibility, making it a great candidate for bio-electronic and bio-medical applications. We focus on three different DNA sequences where GC base pairs are separated by AT base pairs. Then, we conceptualize a setup where DNA molecules lay horizontally on a substrate and make electrical contact between the central AT base pairs and the underlying gold atoms. Our motivation for horizontal DNA molecules comes from DNA origami tiles which can self-assemble on a surface and form long, repeating DNA wires,[24–27] which could be used in a circuit.

First, we model DNA molecules on top of the Au (111) surface via molecular dynamics (MD) simulations which have been proven to be an effective tool to investigate the DNA-gold interactions.[28,29] Our calculations agree with prior work which finds several different conformations due to interaction with the gold surface. Then, we cluster the resulting conformations and use the top ten groups for the quantum mechanical calculations. We employ density functional theory (DFT) calculations to study the electronic properties of each selected DNA conformation. We present that the energy and spatial distribution of the molecular orbitals vary with sequence and molecular conformation. Next, we use Green's function method to perform charge transport calculations by considering the alterations in contact atoms caused by conformational changes. Our results demonstrate that the transmission of the HOMO energy varies by up to 60 times between different conformations.

This paper is organized as follows: first, we examine the structural changes of DNA lying on top of an Au (111) substrate. Subsequently, we present a comprehensive analysis of differences in conformations of each DNA sequence. This is followed by an in-depth investigation of the energy levels and molecular orbitals of each conformation. Then, we report the charge transport calculations through DNA aiming to elucidate the effect of the sequence and conformation on its electrical properties.

## Results and Discussion

The interaction of DNA and solid substrates is influenced by many factors such as the crystal structure of surfaces, orientation of atoms at the surface, the presence of surface roughness and/or defects, the concentration, size, and sequence of the DNA molecule, and environmental conditions. It's challenging to account for all these factors and both model and understand how DNA surface interactions affect conformation. To gain an understanding of how electronic properties depend on conformation both in a single sequence and between closely related sequences, we study three sequences that consist of the same content, but the position of the nucleobases differs: 1) ss($A_3$)-ds($C_3T_3C_3$)-ss($A_3$), 2) ss($A_3$)-ds($C_3A_3C_3$)-ss($A_3$), and 3) ss($A_3$)-ds($G_3A_3G_3$)-ss($A_3$). We refer to these sequences as $C_3T_3C_3$, $C_3A_3C_3$, and $G_3A_3G_3$, respectively. We add a single-strand poly adenine extension to each 9 base-pair long double-stranded DNAs for better adsorption to the Au (111) surface as illustrated in Figure 1. We restrict the length of the sequences to nine base pairs due to the limitations of DFT calculations.

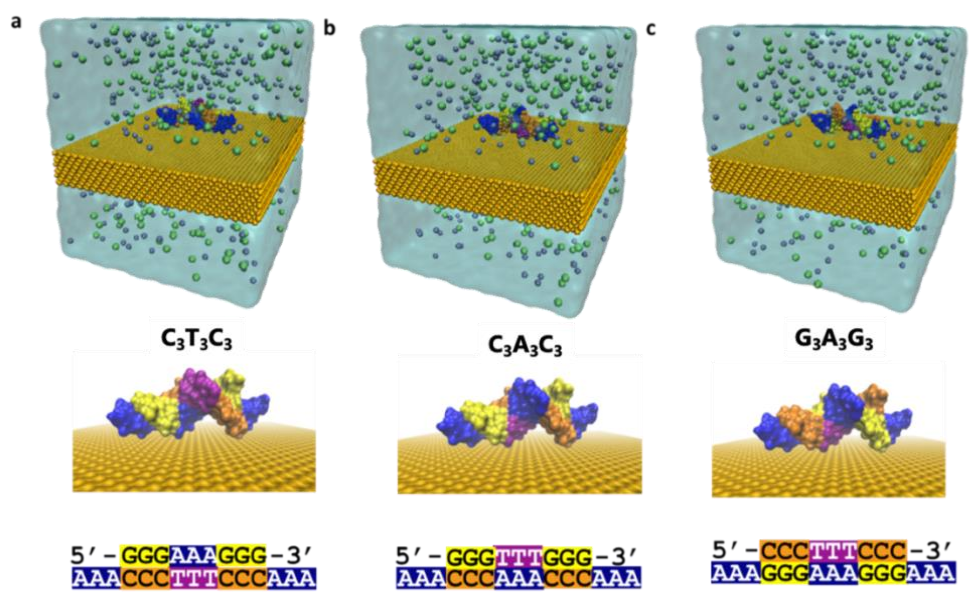

**Figure 1:** Simulation system a) for ss($A_3$)-ds($C_3T_3C_3$)-ss($A_3$) named as $C_3T_3C_3$, b) ss($A_3$)-ds($C_3A_3C_3$)-ss($A_3$) named as $C_3A_3C_3$, c) ss($A_3$)-ds($G_3A_3G_3$)-ss($A_3$) named as $G_3A_3G_3$. The top figures show the simulation box including water and 0.15 KCl ions (gray: K & green: Cl). Below, a side view of DNA on top of Au (111) is shown without water molecules and ions for clarity. DNA sequences are given at the bottom part for each case.

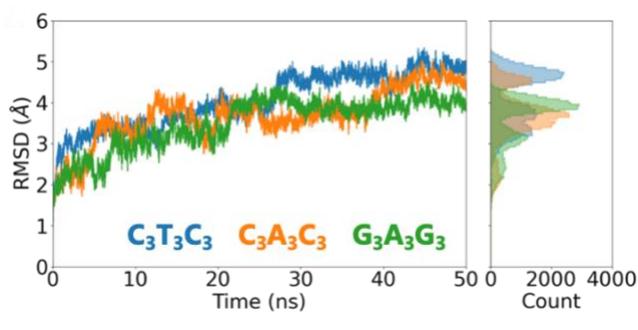

**Figure 2:** RMSD change within 50 ns of MD simulation together with RMSD histogram which shows RMSD mean and variation for each sequence.

To reduce the computational time required for the MD simulations, we place the DNA structures 3.5 Å away from the surface. Then we solvate the whole system with a 0.15 M KCl. We use the TIP3P forcefield[30] to model water molecules and CHARMM36[31] and INTERFACE[32] force fields for the DNA and gold substrate, respectively. It's known that when gold surface is in contact with water, surfactants, and biopolymers, an attractive polarization takes place at the interface[33] due to highly mobile electrons. Therefore, while modeling the interactions between DNA and gold surfaces, it is important to include polarization effects on the surfaces. We use the INTERFACE force field, which has been generated using experimental material properties like density and surface energy as target data to fit the Lennard Jones (LJ) atomic radii and well depth, respectively.[32] Consequently, in our simulations we only include Van der Waals forces, and we neglect further Coulomb interactions between the DNA molecules and the gold substrate. We simulate the DNA - Au (111) systems for 50 ns in the NPT ensemble using the NAMD[34] program. During the simulations, we keep the location of the Au atoms fixed and neglect the reconstruction of the surface for simplicity. We provide a comprehensive description of the simulation procedure in the Method section in the SI.

We first assess the global conformational changes of DNA upon placement on the surface using the conventional measure of the root-mean-square deviation (RMSD). The RMSD increases during the first 20 ns of the trajectory before reaching a plateau for $C_3T_3C_3$ and $G_3A_3G_3$ while it takes almost 40 ns for $C_3A_3C_3$ to reach a plateau as shown in Figure 2. We observe that the final RMSD values converged to ~5 Å indicating that significant structural changes occurred for all three DNA sequences considered. Thereafter, we focus on the Root Mean Square Fluctuation (RMSF) for each individual atom to reveal the regions of the DNA structures that are most flexible (Figure S2). RMSF plots show that the DNA molecules exhibit significant mobility in the terminal region of the structure in all cases. On the other hand, the central AT region for each case stays stable during the 50 ns simulation. Therefore, we anticipate that with an increase in simulation time, the terminal regions will continue to fluctuate, causing more conformational changes to the molecule, while the central AT-rich region will remain stable for an extended period. Since our focus is on the electronic properties of DNA and understanding the effects of conformational changes, rather than identifying the most thermodynamically stable conformation of DNA on a gold substrate, we restrict the simulation time to 50 ns. However, we acknowledge that the conformation of such short DNA sequences may exhibit greater conformational changes when the simulation time is extended.

Next, we calculate pairwise RMSD plots to monitor the conformational changes with time for both (i) the whole DNA structure and (ii) the central AT region. Figure S3 illustrates the temporal evolution of the molecule's conformation by comparing the RMSD change between the conformations saved every 1 ns. The right column of Figure S3 shows that the central AT-region is stable and displays minimal conformational variability. We also observe the stability of this region in the RMSF plots as previously mentioned. Conversely, the terminal regions cause the whole structure to exhibit clusters of distinct conformations (see left column of Figure S3). Therefore, we use a RMSD-based clustering algorithm[35] and categorize the conformations of the whole DNA with a cutoff value of 1.5 Å RMSD. Then, we select the top ten clustered groups for further analysis. We consider the conformations that don't fit any of the top ten groups as un-clustered. Table S1 presents the number of conformations and their percentages for each cluster.

It is important to quantify the nature of the interaction between the DNA strand and substrate as this results in the various transmission/conductance features which is also previously reported in the literature.[15,36,37] From each of the ten different clusters, we choose the conformation that has the minimum RMSD difference from all others in the same cluster – we call this *the representative structure*. Because we only take a snapshot from MD simulations, depending on where we are in the conformational space, we may end up with a conformation that is slightly away from the local energy minimum. Therefore, we apply energy minimization to all thirty representative structures before DFT calculations. Figure S4 illustrates all these structures. To identify the conformational differences, we initially calculate the percentage of contact atoms, which were defined as the DNA atoms within 5 Å of the Au (111) surface. Our observations indicate that, in all sequences, at least one nucleobase from each AAA extension is contacting the Au (111) substrate. In the $C_3T_3C_3$ sequence, two of the Adenines are fully adsorbed for most of the representative structures, while in $C_3A_3C_3$ and $G_3A_3G_3$ it decreases to one. Furthermore, we don't observe 100% interaction with the gold substrate in any nucleobase other than AAA extensions. We also note in the complementary strand lacking the AAA extension region, the nucleobases close to the GC-AT boundaries interact with the substrate for each cluster. Although these interactions depend on the initial placement of the DNA molecules on the substrate, our results indicate that the relative positions of nucleobases result in different conformations on the surface.

Subsequently, we compare the hydrogen bonds between the base pairs for each conformation (Figure S5). Notably, while the representative structures displayed various percentages of contact atoms, we see comparable numbers of hydrogen bonds across all representative structures. This indicates that the terminal AAA extensions provide base pair integrity for the dsDNA part of the sequences during the simulation time.

Next, we change our attention towards investigating the puckering angle since, A- and B-form DNAs exhibit distinct conductance values. Figure S6 demonstrates the sugar puckering angle for all representative structures, thereby shedding light on the potential DNA form changes. Interestingly, we note that most of the Ts displays pucker angle close to the A-form, rather than the B-form, while the rest of the structure stays close to the B-form sugar puckering.

In the next section, we study the electronic properties of each representative structure, which reveals that conformational changes induced by substrate interactions play an important role in determining the transmission.

**Effect of Conformations on Electronic Properties**

We remove the Au (111) substrate, water molecules, and counterions from the representative structures for DFT calculations, and we set the total charge of each system to −22, which corresponds to the number of phosphate groups in the backbone. We carry out the DFT calculations with the B3LYP exchange-correlation function and 6-31G(d,p) basis set together with the polarizable continuum model (see details in the Methods section in SI).

First, we examine the effect of conformational changes on the energy levels of the molecular orbitals (MO). Figure S7 illustrates the first 11 orbitals from both occupied (HOMO-10 to HOMO) and unoccupied (LUMO to LUMO+10) states. We note the difference between the highest and lowest HOMO levels is 155 meV for $C_3A_3C_3$, 115 meV for $C_3T_3C_3$, and 153 meV for $G_3A_3G_3$. We find that $C_3A_3C_3$ exhibits the highest HOMO energy difference between the representative structures. Furthermore, we also observe that $C_3A_3C_3$ displays less variation in the energy of the HOMO levels in comparison to the other two sequences.

Furthermore, we observe that HOMO levels are generally higher for $C_3T_3C_3$ and $G_3A_3G_3$: 70% of the $C_3T_3C_3$ and $G_3A_3G_3$ conformations have a HOMO level above −5 eV, while it is only 20% for $C_3A_3C_3$. Moreover, although there are shifts in the orbital energy levels, the band gaps of all representative structures lie within the 3.8 – 4.2 eV range (Figure S8). This clearly indicates that while the content of each sequence is the same, energy levels differ due to conformational changes.

Later, we analyze the spatial distribution of MO along the molecules. Figure S9 represents the first 4 occupied MOs from each representative structure. For all cases, occupied orbitals are either located on the rightmost or leftmost guanines. However, depending on the conformation, the localization of orbitals changes; for instance, while in $C_3A_3C_3$ C1 (Cluster 1), the HOMO and HOMO-4 are delocalized on the same side, in C2 they are on the opposite guanines. As a result, it is reasonable to expect sample-to-sample variations in the transmission values.

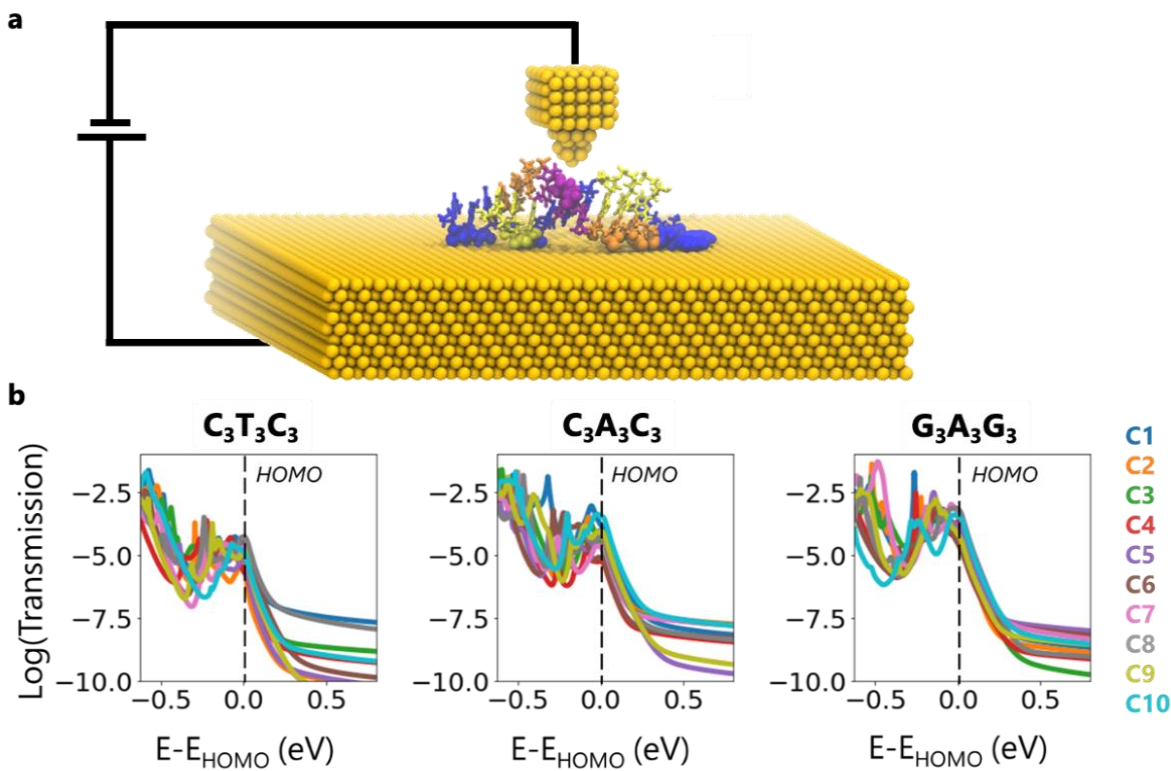

**Figure 3:** a) Schematic representation of the conceptualized calculation setup, note that gold electrodes aren't explicitly included in the DFT calculations b) Transmission plots for every conformation of the $C_3T_3C_3$, $C_3A_3C_3$ and $G_3A_3G_3$ sequences.

To calculate the transmission, we consider a conceptual scenario where the DNA on Au (111) substrate makes a contact from the top of the DNA molecule as in STM and/or AFM measurements (Figure 3a). We specifically place the top contact on the backbone atoms of the central AT base pair to ensure that the most stable portion of the structure is probed. We chose this configuration based on MD simulation results which reveal that the central region is the most stable portion of the structure. Here, the top contact can be thought of as a gold tip (Figure 3a), different interacting molecules, or covalently bonded functionalization groups. As for the bottom contact, we select the contact atoms that we previously defined in Figure S5. It should be noted that as the conformation changes, so does the number of contact atoms (Figure S5, Figure 4).

As mentioned earlier, we do not include explicit Au atoms in the DFT calculations due to computation limitations. Therefore, in modeling the charge transport properties along the DNA molecules, we need open boundary conditions resulting from the top and the bottom contact. To model charge transport including these open boundary conditions, we employ Green's function technique, which includes self-energy terms corresponding to each contact. Details about the model and calculation parameters are given in the Methods section of the SI.

Figure 3b demonstrates the transmission plots for the ten conformations corresponding to each sequence. We shift the energy axis to make the HOMO energy of all representative structures to be at E=0. First, we notice that for a given sequence, the transmission can vary by order of magnitude at the HOMO. The conformations which have the highest and the lowest transmission, as well as the variance between them, depends on the sequence. For instance, at HOMO, while the transmission difference between C4 and C6 of $C_3T_3C_3$ reaches up to 23 times, between C4 and C10 of $C_3A_3C_3$, it is 60 times and between C6 and C8 of $G_3A_3G_3$, it is 24 times. Inside the band gap, these differences reach up to 3 orders of magnitude due to the broadening of energy levels near the HOMO/LUMO edges, which increases the density of states inside the bandgap. Furthermore, we observe that the highest transmission does not always correspond to C1, which is the most observed conformation in the simulations.

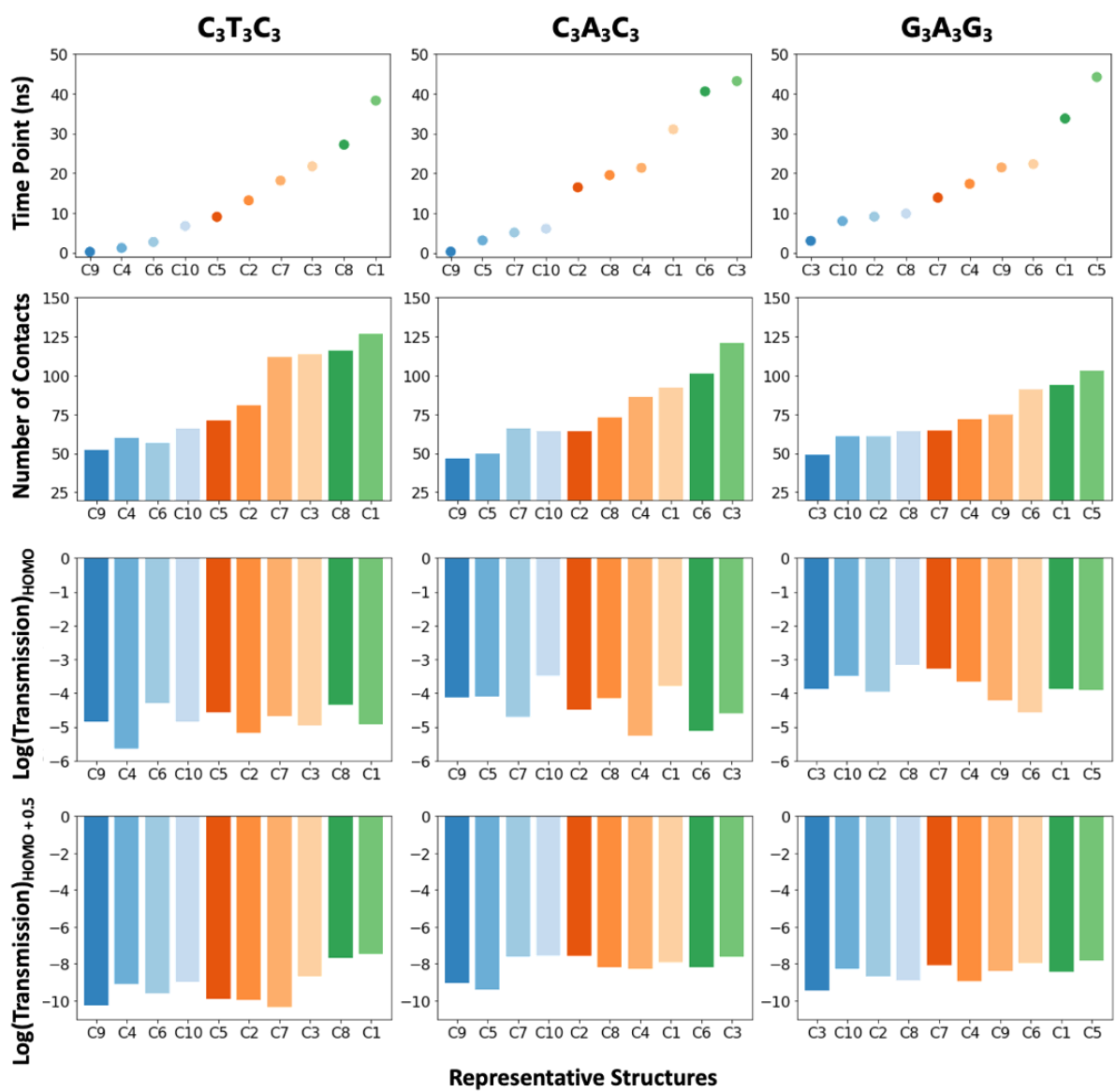

**Figure 4:** Bar plots showing the relation between time points, number of contacts and transmission at the HOMO and HOMO+0.5 for the $C_3T_3C_3$, $C_3A_3C_3$ and $G_3A_3G_3$ sequences.

Next, we analyze the relation between each representative structure and the number of contacts, transmission at the HOMO and the band gap by plotting the results based on time intervals that the representative structures correspond to (Figure 4). The transmission at the band gap is selected from 0.5 eV away from the HOMO. The bar plots given in Figure 4 indicate that the number of contacts increases as the simulation progresses in time, and this influences the transmission at the band gap. On the other hand, the transmission at the HOMO doesn't demonstrate one to one correlation with the number of contacts. We think this is related to the spatial distribution of the HOMOs (Figure S9), i.e., the HOMO transmission values are highly dominated by the HOMO only, but the transmission values at the band gap is the collective effect of the other molecular orbitals as well.

All the results indicate that changing the position of base pairs affects 1) adsorption onto a gold substrate, 2) conformation, 3) the number of contact atoms, 4) energy levels and spatial locations of the molecular orbitals. The cumulative effects of all these change the transmission of DNA molecules. While designing a DNA nanowire that is going to be used with a gold substrate, it's important to consider all these effects. Our results also indicate that it's challenging to predict transmission values just with sequence selection, since there will be different conformations.

**Limitations of the models**

Although we present significant results for the community, we acknowledge that our study has some limitations. First, we may not have selected the adsorbed geometries from the whole conformational ensemble. Because of the high processing demands, longer (microseconds or milliseconds) simulations for various initial structures are hard to achieve, resulting in insufficient sampling of conformations. Second, we neglect the surface changes of gold substrate and the presence of the top contacts which can also affect the final conformations and transmission. Finally, due to convergence issues, we didn't include the gold atoms explicitly in electronic property calculations which may affect the relative conductance values. Each of these issues are difficult standalone challenges that are worthy of further exploration.

## Conclusions

Many DNA-based electronic device applications rely on understanding how the molecules interact with substrates. Critical aspects include the durability of DNA on surfaces and the electron pathways flowing from contacts to the DNA and finally into the substrate. There are many factors influencing the interaction of DNA with solid substrates such as the substrate's crystal structure, surface roughness and/or defects, the sequence, structure, and size of the DNA molecules, organization on the surface, and environmental conditions. In this paper, as an initial step to understanding the role of these factors in determining both the conformation and conductance, we studied a model system composed of a nine-base pair DNA extended with three single-strand Adenines at each end to provide adsorbability. To investigate the influence of interaction with a substrate on charge transmission of DNA, we brought the DNA molecules to interact with the Au (111) substrate in an explicit solvent environment. Our MD simulations revealed DNA would be significantly changed its structure to adsorb on gold substrate, especially from the extension (A3) regions. This leads us to encounter several different conformations instead of one. We found that all adenine extensions weren't fully adsorbed onto the substrate after the 50 ns simulation. Although the sequence content was the same for all three cases, changing the relative positioning of nucleobases alters the contact points (i.e., the DNA atoms interacting with the Au substrate). Our ab initio calculations, performed on the top ten conformations of each sequence ($C_3T_3C_3$, $C_3A_3C_3$, $G_3A_3G_3$), revealed that between different conformations both energy levels and spatial locations of molecular orbitals vary. In our charge transport model, interacting atoms with the gold substrate served as the bottom electrical contact, and backbone atoms at the center triplet of the molecule served as the second contact, which mimics a weakly interacting conducting AFM or STM tip. With Green's function-based charge transmission calculations, we found that at the HOMO the transmission can differ up to 60 times among different conformations. As the number of contact points increases, the transmission at the HOMO is not directly affected. Conversely, the transmission at the band gap region appears to be dependent on the number of contact atoms. Overall, we demonstrated that the interaction between DNA and gold substrate is important for charge transmission and needs to be considered while developing DNA-based electronic devices. The knowledge to achieve measurable/differentiable charge transmission values obtained in this paper can pave the way for designing new DNA-based electronics such as sensors and sequencing devices.

## Author Contributions

We strongly encourage authors to include author contributions and recommend using CRediT for standardised contribution descriptions. Please refer to our general author guidelines for more information about authorship.

## Conflicts of interest

There are no conflicts to declare.

## Acknowledgements

We acknowledge NSF ECCS Grant Numbers 1807391 and 1807555. We acknowledge using the Hyak supercomputer system at the University of Washington and TUBITAK ULAKBIM, High Performance and Grid Computing Center (TRUBA resources). Busra Demir further acknowledges a TUBITAK 2214-A International Doctoral Research Fellowship.

## References

1. N. C. Seeman and H. F. Sleiman, DNA nanotechnology, *Nat. Rev. Mater.*, 2018, **3**, 17068.
2. S. Jäger, G. Rasched, H. Kornreich-Leshem, M. Engeser, O. Thum and M. Famulok, A Versatile Toolbox for Variable DNA Functionalization at High Density, *J. Am. Chem. Soc.*, 2005, **127**, 15071–15082.
3. R. G. Endres, D. L. Cox and R. R. P. Singh, Colloquium: The quest for high-conductance DNA, *Rev. Mod. Phys.*, 2004, **76**, 195–214.
4. J. C. Genereux and J. K. Barton, Mechanisms for DNA charge transport, *Chem. Rev.*, 2010, **110**, 1642–1662.
5. S. O. Kelley and J. K. Barton, Electron transfer between bases in double helical DNA, *Science*, 1999, **283**, 375–381.
6. L. Xiang, J. L. Palma, C. Bruot, V. Mujica, M. A. Ratner and N. Tao, Intermediate tunnelling–hopping regime in DNA charge transport, *Nat. Chem.*, 2015, **7**, 221–226.
7. Y. Li, L. Xiang, J. L. Palma, Y. Asai and N. Tao, Thermoelectric effect and its dependence on molecular length and sequence in single DNA molecules, *Nat. Commun.*, 2016, **7**, 1–8.
8. C. Bruot, L. Xiang, J. L. Palma and N. Tao, Effect of Mechanical Stretching on DNA Conductance, *ACS Nano*, 2015, **9**, 88–94.
9. L. Xiang, J. L. Palma, Y. Li, V. Mujica, M. A. Ratner and N. Tao, Gate-controlled conductance switching in DNA, *Nat. Commun.*, 2017, **8**, 1–10.
10. R. Zhuravel, H. Huang, G. Polycarpou, S. Polydorides, P. Motamarri, L. Katrivas, D. Rotem, J. Sperling, L. A. Zotti, A. B. Kotlyar, J. C. Cuevas, V. Gavini, S. S. Skourtis and D. Porath, Backbone charge transport in double-stranded DNA, *Nat. Nanotech.*, 2020, **15**, 836–840.
11. H. Mehrez and M. P. Anantram, Interbase electronic coupling for transport through DNA, *Phys. Rev. B.*, 2005, **71**, 115405.
12. S. R. Patil, H. Mohammad, V. Chawda, N. Sinha, R. K. Singh, J. Qi and M. P. Anantram, Quantum Transport in DNA Heterostructures: Implications for Nanoelectronics, *ACS Appl. Nano Mater.*, 2021, **4**, 10029–10037.
13. H. Mohammad, B. Demir, C. Akin, B. Luan, J. Hihath, E. E. Oren and M. P. Anantram, Role of intercalation in the electrical properties of nucleic acids for use in molecular electronics, *Nanoscale Horiz.*, 2021, **6**, 651–660.
14. S. Bag, S. Mogurampelly, W. A. Goddard and P. K. Maiti, Dramatic changes in DNA conductance with stretching: Structural polymorphism at a critical extension, *Nanoscale*, 2016, **8**, 16044–16052.
15. A. Aggarwal, A. K. Sahoo, S. Bag, V. Kaliginedi, M. Jain and P. K. Maiti, Fine-tuning the DNA conductance by intercalation of drug molecules, *Phys. Rev. E*, 2021, **103**, 032411.
16. T. Harashima, C. Kojima, S. Fujii, M. Kiguchi and T. Nishino, Single-molecule conductance of DNA gated and ungated by DNA-binding molecules, *Chem. Commun.*, 2017, **53**, 10378–10381.
17. C. Guo, K. Wang, E. Zerah-Harush, J. Hamill, B. Wang, Y. Dubi and B. Xu, Molecular rectifier composed of DNA with high rectification ratio enabled by intercalation, *Nat. Chem.*, 2016, **8**, 484–490.
18. J. Qi, N. Edirisinghe, M. G. Rabbani and M. P. Anantram, Unified model for conductance through DNA with the Landauer-Büttiker formalism, *Phys. Rev. B*, 2013, **87**, 085404.
19. L. Xiang, J. L. Palma, C. Bruot, V. Mujica, M. A. Ratner and N. Tao, Intermediate tunnelling–hopping regime in DNA charge transport, *Nat. Chem.*, 2015, **7**, 221–226.
20. B. Giese, Long-Distance Electron Transfer Through DNA, *Annu. Rev. Biochem.*, 2002, **71**, 51–70.
21. P. B. Woiczikowski, T. Kuba, R. Gutírrez, R. A. Caetano, G. Cuniberti and M. Elstner, Combined density functional theory and Landauer approach for hole transfer in DNA along classical molecular dynamics trajectories, *J. Chem. Phys.*, 2009, **130**, 215104.
22. J. M. Artés, Y. Li, J. Qi, M. P. Anantram and J. Hihath, Conformational gating of DNA conductance, *Nat. Commun.*, 2015, **6**, 8870.
23. C. Bruot, L. Xiang, J. L. Palma and N. Tao, Effect of mechanical stretching on DNA conductance, *ACS Nano*, 2015, **9**, 88–94.
24. G. Tikhomirov, P. Petersen and L. Qian, Triangular DNA Origami Tilings, *J. Am. Chem. Soc.*, 2018, **140**, 17361–17364.
25. G. Tikhomirov, P. Petersen and L. Qian, Fractal assembly of micrometre-scale DNA origami arrays with arbitrary patterns, *Nature*, 2017, **552**, 67–71.
26. D. Liu, S. H. Park, J. H. Reif and T. H. LaBean, DNA nanotubes self-assembled from triple-crossover tiles as templates for conductive nanowires, *Proc. Natl. Acad. Sci. U S A*, 2004, **101**, 717.
27. S. H. Park, C. Pistol, S. J. Ahn, J. H. Reif, A. R. Lebeck, C. Dwyer and T. H. LaBean, Finite-Size, Fully Addressable DNA Tile Lattices Formed by Hierarchical Assembly Procedures, *Ange. Chem. Int. Edt.*, 2006, **45**, 735–739.
28. O. S. Lee and G. C. Schatz, Molecular dynamics simulation of DNA-functionalized gold nanoparticles, *J. Phys. Chem. C*, 2009, **113**, 2316–2321.
29. C. Izanloo, Evaluation of effect of functionalized gold nanoparticles with a partial negative charge on stability of DNA molecule: a study of molecular dynamics simulation, *Struct. Chem.*, 2018, **29**, 1417–1425.
30. W. L. Jorgensen, J. Chandrasekhar, J. D. Madura, R. W. Impey and M. L. Klein, Comparison of simple potential functions for simulating liquid water, *J. Chem. Phys.*, 1983, **79**, 926–935.
31. J. Huang and A. D. Mackerell, CHARMM36 all-atom additive protein force field: Validation based on comparison to NMR data, *J. Comput. Chem.*, 2013, **34**, 2135–2145.
32. H. Heinz, T. J. Lin, R. Kishore Mishra and F. S. Emami, Thermodynamically consistent force fields for the assembly of inorganic, organic, and biological nanostructures: The INTERFACE force field, *Langmuir*, 2013, **29**, 1754–1765.
33  H. Heinz, K. C. Jha, J. Luettmer-Strathmann, B. L. Farmer and R. R. Naik, Polarization at metal-biomolecular interfaces in solution, *J. R. Soc. Interface*, 2011, **8**, 220–232.
34  J. C. Phillips, D. J. Hardy, J. D. C. Maia, J. E. Stone, J. v. Ribeiro, R. C. Bernardi, R. Buch, G. Fiorin, J. Hénin, W. Jiang, R. McGreevy, M. C. R. Melo, B. K. Radak, R. D. Skeel, A. Singharoy, Y. Wang, B. Roux, A. Aksimentiev, Z. Luthey-Schulten, L. v. Kalé, K. Schulten, C. Chipot and E. Tajkhorshid, Scalable molecular dynamics on CPU and GPU architectures with NAMD, *J. Chem. Phys.*, 2020, **153**, 044130.
35  W. Humphrey, A. Dalke and K. Schulten, VMD: Visual molecular dynamics, *J. Mol. Graph*, 1996, **14**, 33–38.
36  R. Gutiérrez, R. A. Caetano, B. P. Woiczikowski, T. Kubar, M. Elstner and G. Cuniberti, Charge transport through biomolecular wires in a solvent: Bridging molecular dynamics and model hamiltonian approaches, *Phys. Rev. Lett.*, 2009, **102**, 208102.
37  A. Troisi, A. Nitzan and M. A. Ratner, A rate constant expression for charge transfer through fluctuating bridges, *J. Chem. Phys.*, 2003, **119**, 5782.

# SUPPLEMENTARY INFORMATION: DNA—Au (111) Interactions and Transverse Charge Transport Properties for DNA-Based Electronic Devices

Busra Demir[1,2,3,‡], Hashem Mohammad[4,‡], M. P. Anantram[3] and Ersin Emre Oren[1,2,*]

[1]Department of Materials Science & Nanotechnology Engineering, TOBB University of Economics and Technology, Ankara, Turkiye

[2]Bionanodesign Laboratory, Department of Biomedical Engineering, TOBB University of Economics and Technology, Ankara, Turkiye

[3]Department of Electrical and Computer Engineering, University of Washington, 98195 Seattle, WA, USA,

[4]Department of Electrical Engineering, Kuwait University, P.O. Box 5969, Safat 13060, Kuwait.

[‡] these authors contributed equally

[*] corresponding author

## METHODS

First, we use MD simulations and clustering methods to obtain representative structures, then we use DFT calculations to generate the system Hamiltonian that contains molecular orbital energy level information and their coupling between one another. In the third step, we use the Hamiltonian with the Green's function method to calculate the transverse transmission.

Schematic representations of the modeling steps are shown in Figure S1.

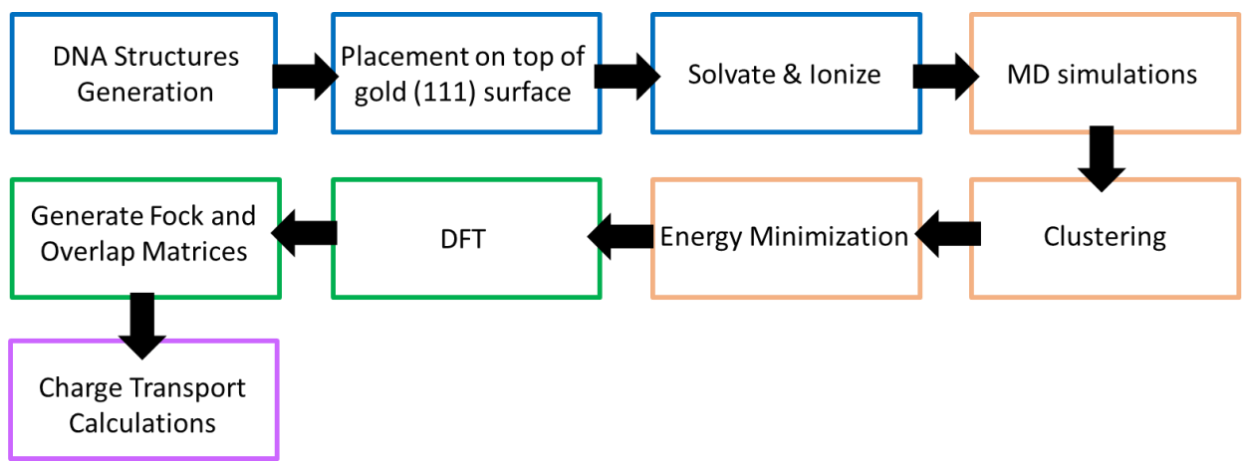

**Figure S1:** Modeling steps used in this study.

# 1. Structure preparation and MD Simulations

We used NAB Tool implemented in AMBER[S1] and VMD's[S2] Inorganic Builder plugin to generate the DNA structures and Au (111) surface respectively. Then, we placed the DNA structures 3.5 Å away from the Au (111) surface and we solvated the system with TIP3P[S3] water molecules and 0.15 M KCl. CHARMM36[S4] and INTERFACE[S5] force fields are used for the DNA and gold substrate respectively. We used 12 Å cutoff to calculate Van der Waals potential energies and the particle-mesh Ewald (PME) method with a maximum grid spacing of 1.5 Å to compute electrostatic interactions. We use Langevin dynamics, and the simulation time-step is set to 1 fs. For all simulation steps, Au atoms were kept fixed. We first minimize the water molecules and the ions for 2000 steps at 295 K while the entire DNA molecule was kept fixed in a constant volume. After the minimization, we first let extended poly-A parts fluctuate for 2 ns while keeping the ds-DNA part of the molecule fixed. Then the entire system is equilibrated for 1 ns while only the gold substrate is fixed. Finally, the production simulations were performed for 50 ns using the Langevin piston Nose-Hoover method implemented in NAMD[S6] to maintain atmospheric pressure. The MD simulation analysis is carried out with pytraj library, VMD plugins and tcl scripting.

The representative structures from the MD simulation were determined by a RMSD-based clustering algorithm within VMD software with 1.5 Å cutoff value as mentioned in the main manuscript. Representative structures were subjected to 5000 steps of energy minimization prior to DFT calculations.

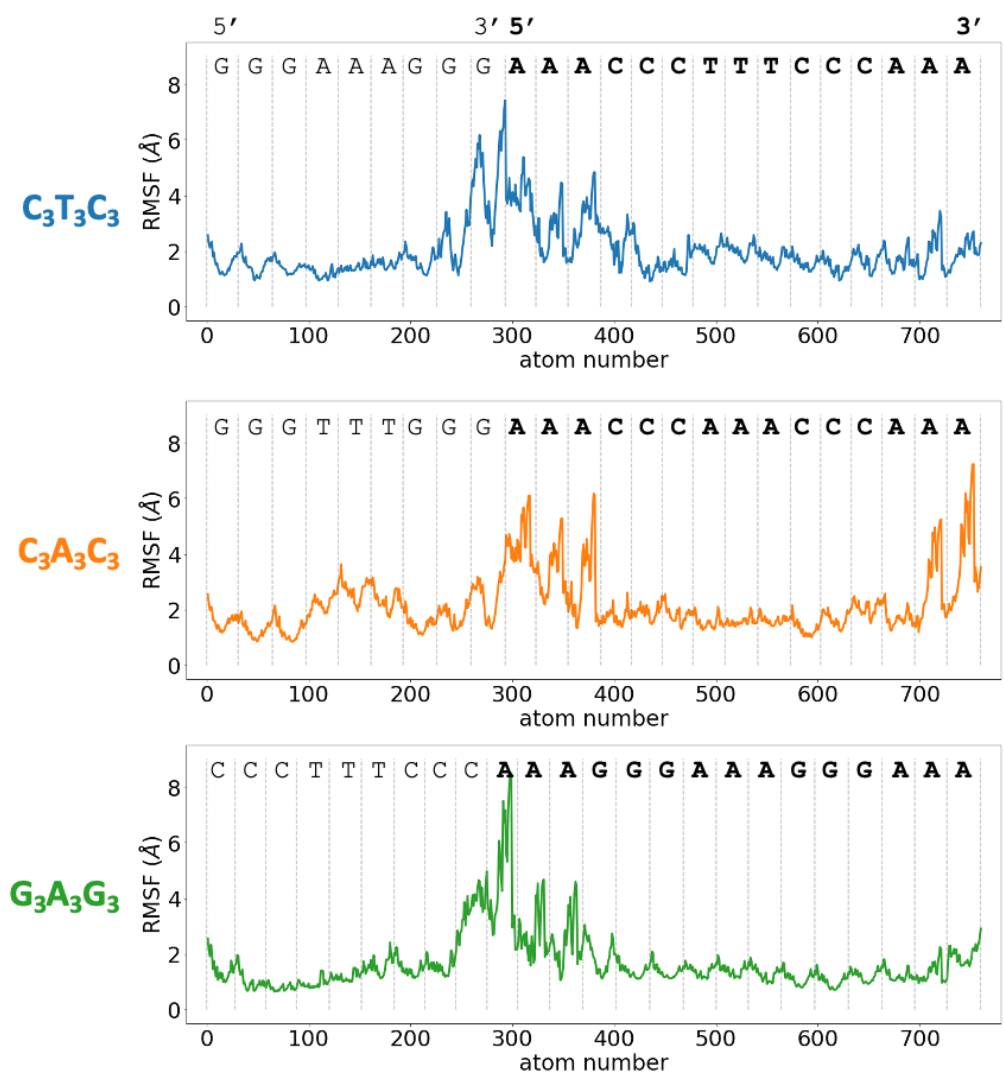

**Figure S2:** Plots of RMSF vs atom number for each sequence. The graph indicates the variation of RMSF for each individual atom in the molecule, denoted by its atom number. The corresponding nucleobases are indicated above each graph. The higher RMSF values indicate greater flexibility and lower RMSF values indicate greater stability.

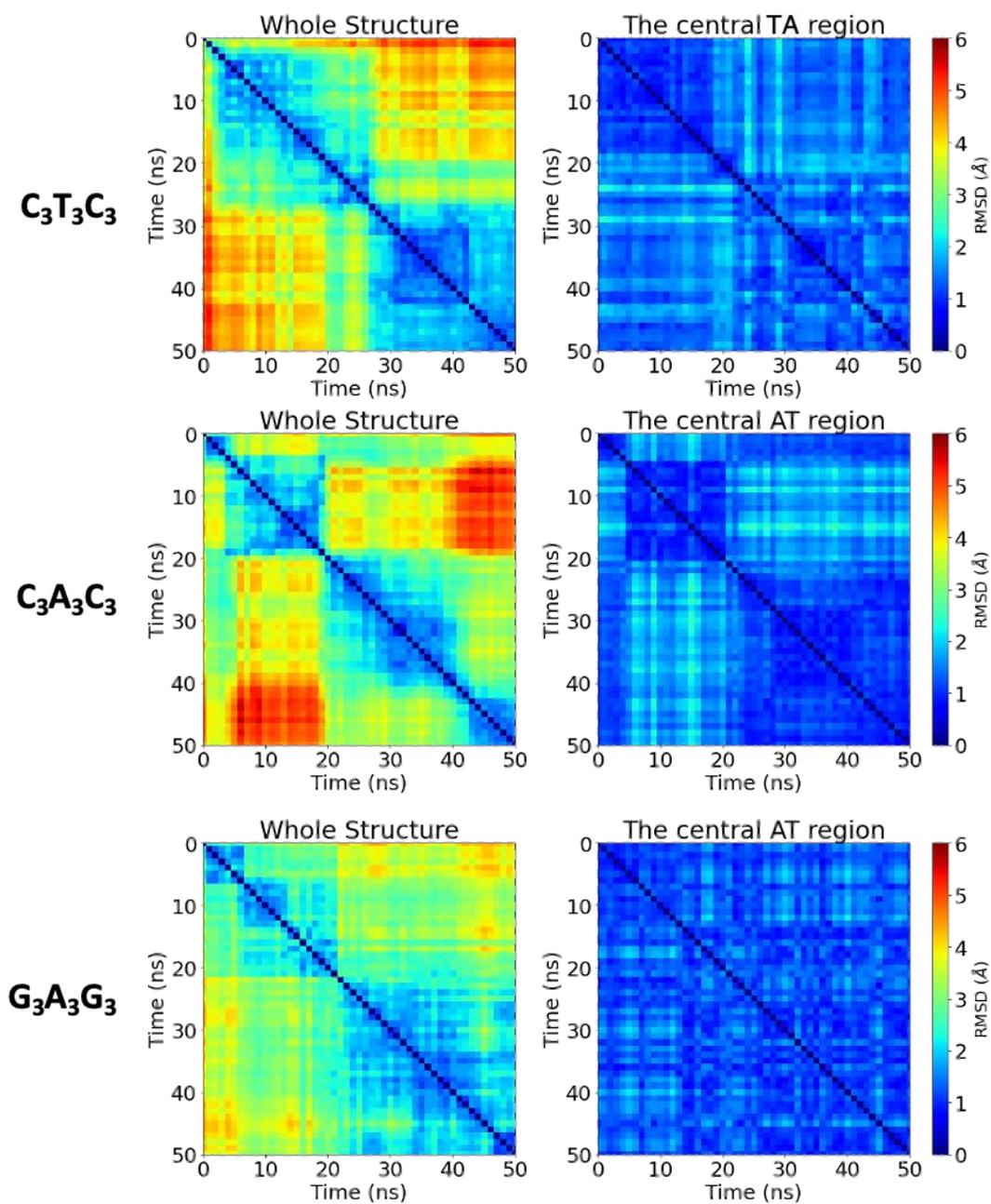

**Figure S3:** Pairwise RMSD analysis for the whole structure and the central triplet region. The graph indicates the variation of RMSD between every conformation saved at 1 ns time interval. Higher values of RMSD indicate a greater structural change between the two conformations.

**Table S1:** Cluster sizes and their percentages

|  | CTC | | CAC | | GAG | |
|---|---|---|---|---|---|---|
|  | # of structures | percentage | # of structures | percentage | # of structures | percentage |
| C1 | 22592 | 45.183% | 15033 | 30.065% | 23244 | 46.487% |
| C2 | 13179 | 26.357% | 13281 | 26.561% | 8478 | 16.956% |
| C3 | 8613 | 17.226% | 7706 | 15.412% | 5256 | 10.512% |
| C4 | 1552 | 3.104% | 4442 | 8.884% | 4737 | 9.474% |
| C5 | 1086 | 2.172% | 4182 | 8.364% | 2769 | 5.538% |
| C6 | 1034 | 2.068% | 2577 | 5.154% | 2070 | 4.140% |
| C7 | 501 | 1.002% | 905 | 1.810% | 1253 | 2.506% |
| C8 | 393 | 0.786% | 801 | 1.602% | 707 | 1.414% |
| C9 | 374 | 0.748% | 390 | 0.780% | 585 | 1.170% |
| C10 | 234 | 0.468% | 99 | 0.198% | 243 | 0.486% |
| Unclustered | 443 | 0.886% | 585 | 1.170% | 659 | 1.318% |

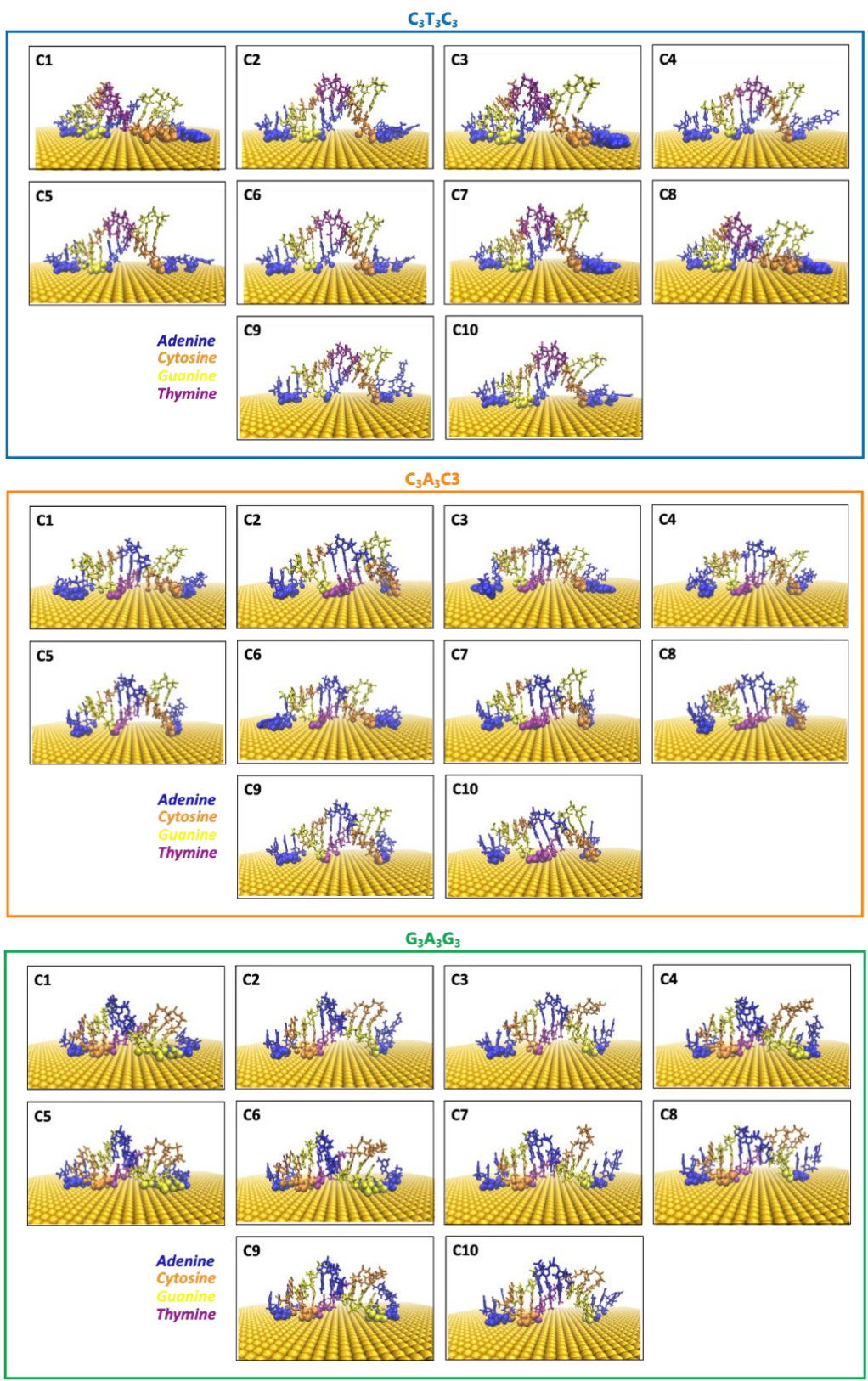

**Figure S4:** Representative structures for all sequences, surface interacting atoms showed as a ball shape. Blue represents Adenines, orange represents cytosine, yellow represents Guanines and pink represents Thymines.

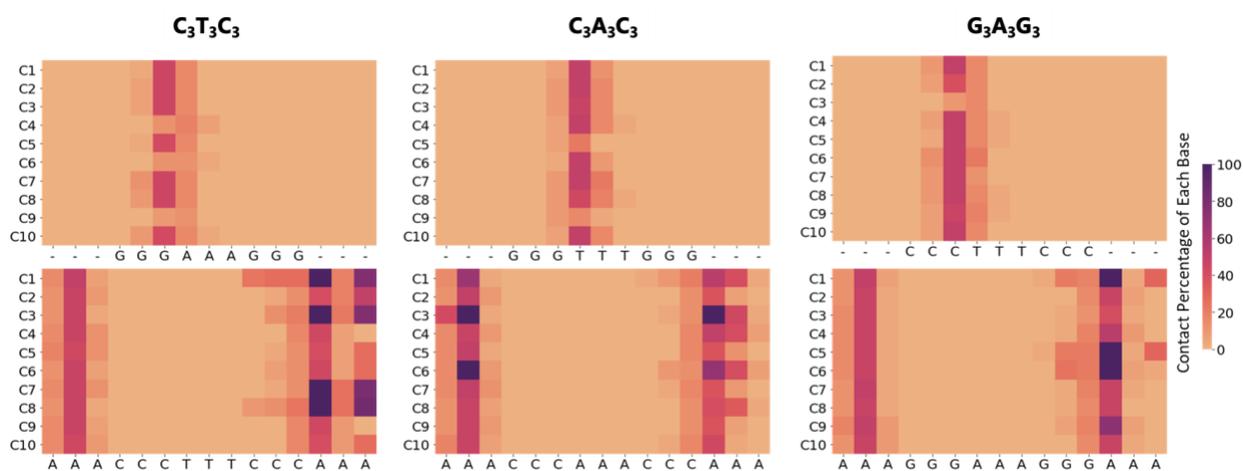

**Figure S5:** Heatmap of the percentage of gold contacting atoms for each representative conformation in CTC, CAC, and GAG cases. The colour gradient represents the percentage of contact atoms in each DNA nucleobase in separate strands, with purple indicating 100% and orange indicating 0% percentage. Each row corresponds to the representative conformation selected from MD simulations, and each column represents residues.

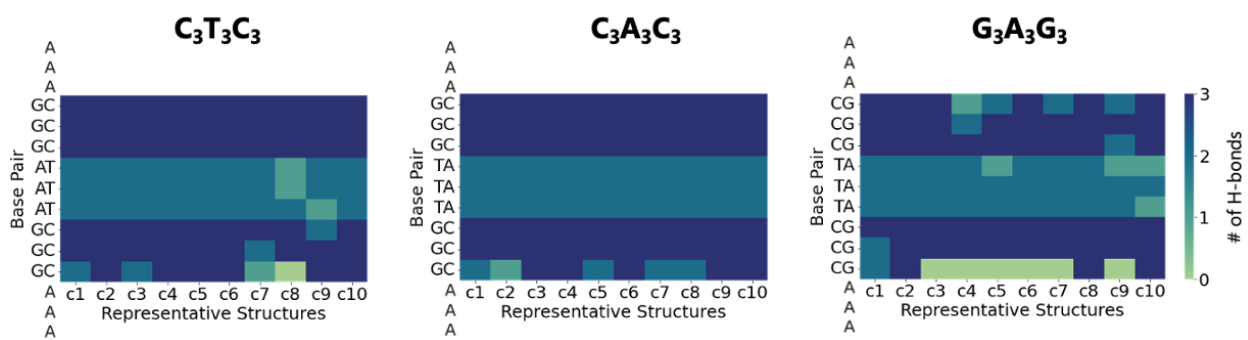

**Figure S5:** Number of hydrogen bonds for the representative structures of each sequence which corresponds to center of the each cluster as described in the main text.

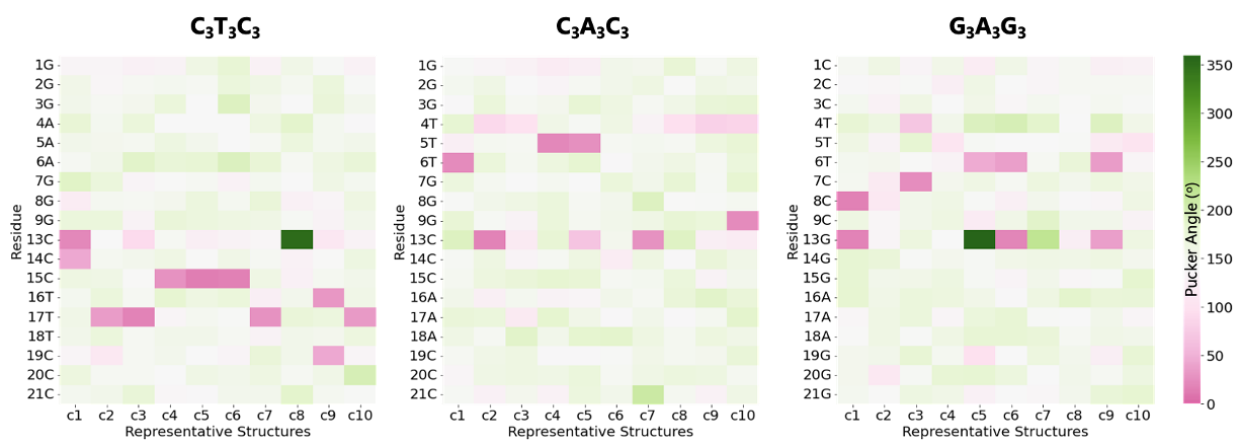

**Figure S6:** Pucker angle analysis for all residues in each cluster for every sequence. The graph indicates the variation of sugar puckering angle. Ideally, the values between 0 and 36° correspond to A-form DNA and 144 and 180° correspond to B-form DNA.

## 2. DFT Calculations

We removed the water molecules, ions, and the gold substrate from the system to enable the convergence of DFT calculations, which were carried out using Gaussian 16[S7] with B3LYP/6-31G(d,p) basis set. We used the polarizable continuum model (PCM) with a dielectric constant of 78.36 to represent the water solvent effect. The total charge of the system was set to the number of phosphate groups in the DNA molecule, which was -22 (the terminal bases do not have any phosphate groups). After achieving convergence, the Fock and Overlap matrices from the DFT calculations were used for the next step.

## 3. Charge Transport Calculations

The quantum transport calculations were carried out using the Green's function method with decoherence probes added to model phase decoherence[S8]. The probe current at each energy were kept zero meaning that the electron does not gain or lose energy while traversing the system. As mentioned in the main manuscript, we assume the contact locations to be at the contacting atoms which are 5 Å away from the surface and the central triplet's backbone atoms. The contact self-energy to model both Au (111) substrate and top contact was set to 600 meV to represent strong coupling. The decoherence scattering rate was set to 100 meV to mimic large decoherence, with an energy decaying factor of 50 meV, limiting the decoherence effect on the onsite potentials (or energy levels) of the nucleotides.

After obtaining the Fock, $H_0$, and overlap matrices, $S_0$ from DFT calculations, we used the Löwdin transformation to convert $H_0$ into a Hamiltonian, $H_{orthogonal}$, in an orthogonal basis set via the following equation:

$$H_{orthogonal} = S_0^{-\frac{1}{2}} H_0 S_0^{-\frac{1}{2}} \quad (1)$$

Here, the diagonal elements of $H_{orthogonal}$ represent the energy levels at each atomic orbital, and the off-diagonal elements correspond to the coupling between the different atomic orbitals. Then, we partitioned the $H_{orthogonal}$ into nucleotides (except the contact locations) and diagonalized the $H_{orthogonal}$ using the following transformation.

$$H = U^\dagger H_{orthogonal} U \quad (2)$$

Here, each block of $H$ is a nucleotide and the diagonal blocks of $H$ are now diagonal matrices. The diagonal elements of the diagonal blocks represent the eigenvalues of the corresponding nucleotide. The off-diagonal blocks of $H$ represent the hopping parameters between the molecular orbitals of the equivalent nucleotides.

The transverse transmission along the molecule was then calculated using Green's function method. The retarded Green's function ($G^r$) was found by solving the following equation:

$$[E - (H + \Sigma_L + \Sigma_R + \Sigma_B)]G^r = I \tag{3}$$

Where $E$ is the energy level, and $H$ is the Hamiltonian defined in Eq. 2. $\Sigma_{L(R)}$ is the left (right) contact self-energy, which represents the coupling strength of the DNA to the bottom (up) contacts by which charge enters and leaves the DNA. The self-energy of the decoherence probe is defined as $\Sigma_B$, which also represents the coupling strength between the DNA and the decoherence probes.

The self-energy of the contacts is defined as $\Sigma_{L(R)} = -\frac{i}{2}\Gamma_{L(R)}$, where $i$ is the imaginary unit. The decoherence probe self-energy is defined as $\Sigma_B(E) = -\frac{\Gamma_k(E)}{2}$, where $k$ represents the k$^{th}$ energy level, and $\Gamma_k$ represents the coupling strength between the probe and the energy level $k$, which is taken as an energy-dependent parameter as follows:

$$\Gamma_k(E) = \Gamma_B \times \exp\left[-\frac{|E - \epsilon_k|}{\lambda}\right] \tag{4}$$

where $\Gamma_B$ determines the value of the decohrence strength, and $\lambda$ is a decay parameter that determines how quickly the decoherence decays away from an energy level.

The decoherence probes were attached to each nucleotide (backbone + base) excluding the contact atoms (top and bottom), where the total number of decoherence probes is 22 (24 nucleotide – 2 contact groups) in the low-bias region, the current at the $i^{th}$ probe calculated with:

$$I_i = \frac{2e}{h}\sum_{j=1}^{N} T_{ij}(\mu_i - \mu_j), \quad i = 1,2,3,\ldots N \tag{5}$$

where $T_{ij} = \Gamma_i G^r \Gamma_j G^a$ is the transmission probability between the $i^{th}$ and $j^{th}$ probes, and $T_{ij} = G^a = (G^r)^\dagger$ is the advanced Green's function. The net current at each decoherence probe should be zero, this yields $N_b$ independent equations from which the following relation can be derived,

$$\mu_i - \mu_L = \left(\sum_{j=1}^{N_b} W_{ij}^{-1} T_{jR}\right)(\mu_R - \mu_L), \qquad i = 1,2,3,\ldots,N_b \qquad (6)$$

Here, $W_{ij}^{-1}$ is the inverse of $W_{ij} = (1 - R_{ii})\delta_{ij} - T_{ij}(1 - \delta_{ij})$, where $R_{ii}$ is the reflection probability at probe $i$, and is given by $R_{ii} = 1 - \sum_{i \neq j}^{N} T_{ij}$. The currents at the top $I_L$ and bottom $I_R$ contacts are not zero because they are governed by the conservation of electron number, $I_L + I_R = 0$. This yields the equation for the current at the left contact as

$$I_L = \frac{2e}{h} T_{eff}(\mu_L - \mu_R) \qquad (7)$$

Comparing equations 5 to 7 yields an effective transmission term:

$$T_{eff} = T_{LR} + \sum_{i=1}^{N_b}\sum_{j=1}^{N_b} T_{Li} W_{ij}^{-1} T_{jR} \qquad (8)$$

In Eq 8, $T_{LR}$ is the coherent transmission from the top electrode to the bottom electrode. The second term is the decoherence contribution to the transmission via decoherence probes. From Eq 7, the zero bias conductance can be approximated as $G = G_0 T_{eff}$, where the quantum of conductance $G_0$, can be calculated as $G_0 = \frac{2e^2}{h} \approx 7.75 \times 10^{-5} \Omega^{-1}$.

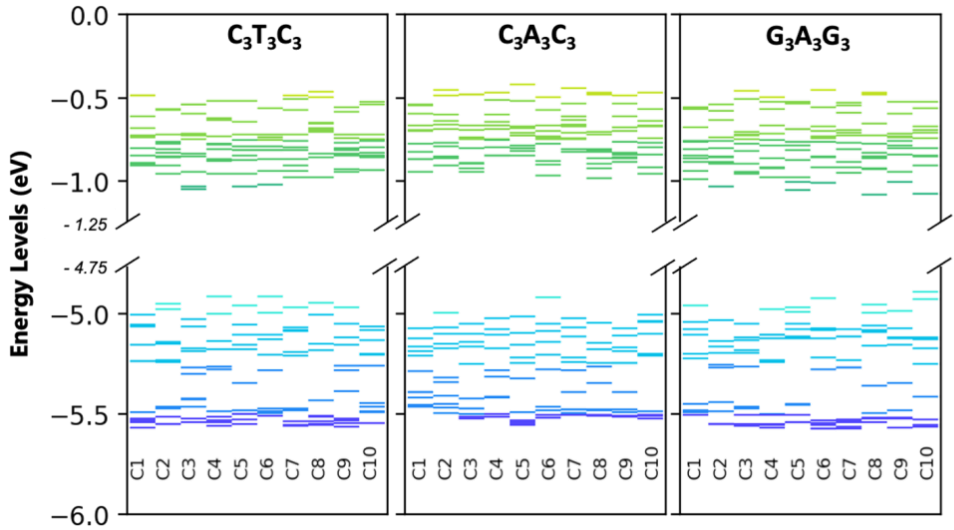

**Figure S7:** Energy band diagram showing only the first 10 orbitals from occupied and unoccupied states for representative structures of $C_3T_3C_3$, $C_3A_3C_3$, and $G_3A_3G_3$ cases.

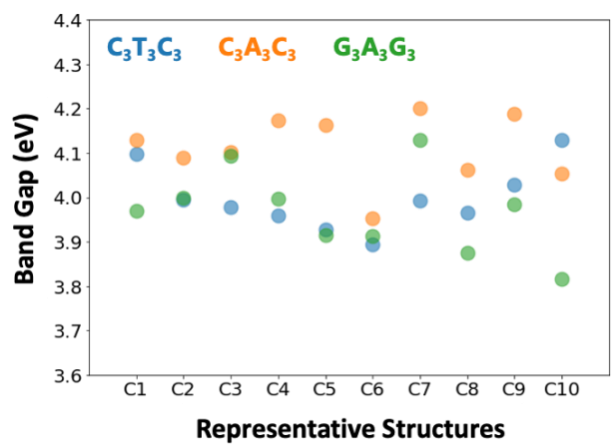

**Figure S8:** Band gap distribution for representative conformations of $C_3T_3C_3$, $C_3A_3C_3$, and $G_3A_3G_3$ cases.

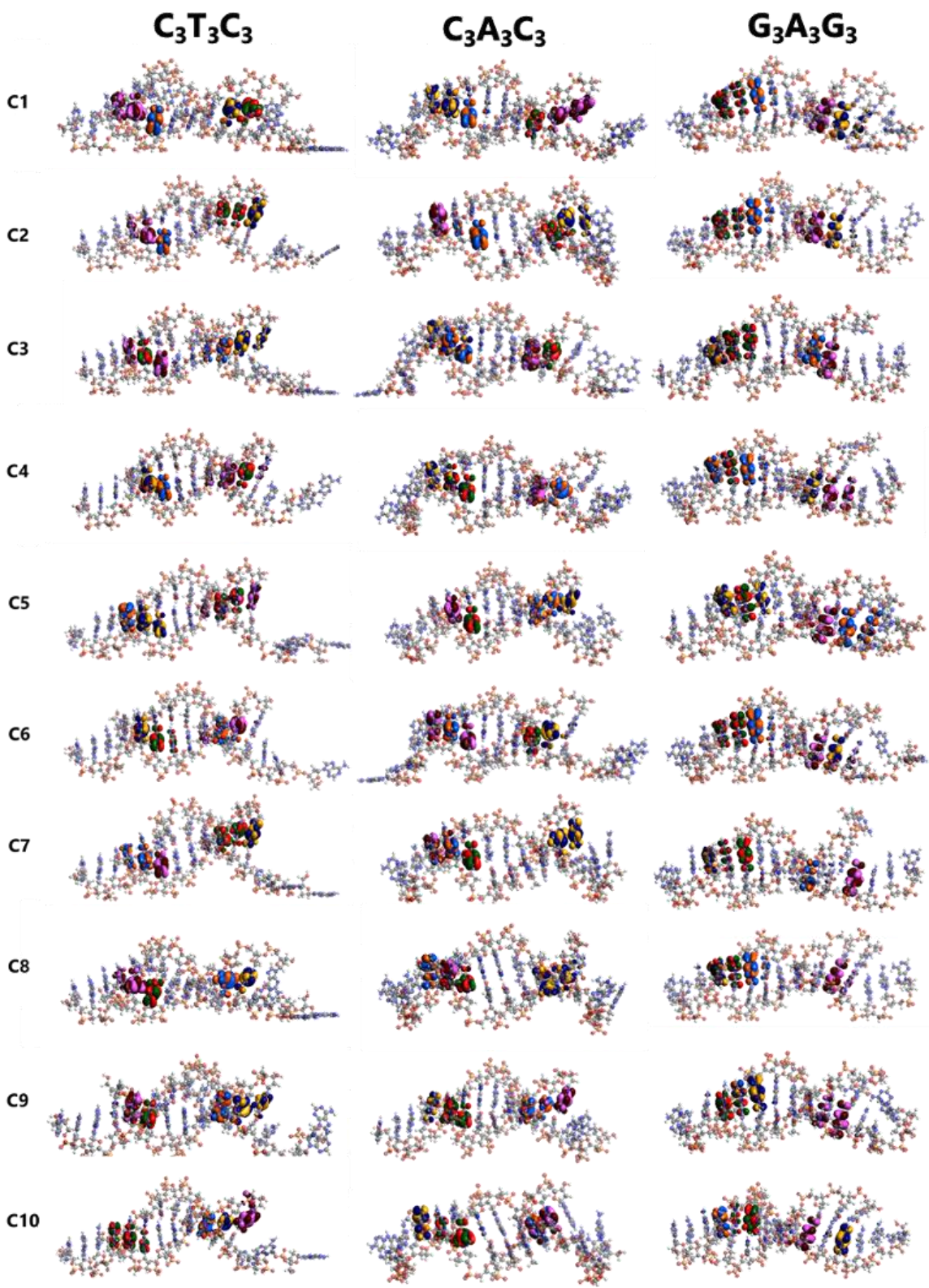

**Figure S9:** Molecular orbitals (for iso value 0.02) for each conformation. Red & Green: HOMO; Blue & Orange: HOMO-1; Yellow & Navy: HOMO-2; Pink & Purple: HOMO-3.

## Supplementary References

S1    D. A. Case, R. M. Betz, D. S. Cerutti, C. I. T.E., T. A. Darden, R. E. Duke, T. J. Giese, H. Gohlke, A. W. Goetz, N. Homeyer, S. Izadi, P. Janowski, J. Kaus, A. Kovalenko, T. S. Lee, S. LeGrand, P. Li, C.Lin, T. Luchko, R. Luo, B. Madej, D. Mermelstein, K. M. Merz, G. Monard, H. Nguyen, H. T. Nguyen, I. Omelyan, A. Onufriev, D. R. Roe, A. Roitberg, C. Sagui, C. L. Simmerling, W. M. Botello-Smith, J. Swails, R. C. Walker, J. Wang, R. M. Wolf, X. Wu, L. Xiao and P. A. Kollman, *AMBER 2016, University of California, San Francisco*.

S2    W. Humphrey, A. Dalke and K. Schulten, *J. Mol. Graph.*, 1996, **14**, 33–38.

S3    W. L. Jorgensen, J. Chandrasekhar, J. D. Madura, R. W. Impey and M. L. Klein, *J. Chem. Phys.*, 1983, **79**, 926–935.

S4    J. Huang and A. D. Mackerell, *J. Comput. Chem.*, 2013, **34**, 2135–2145.

S5    H. Heinz, T. J. Lin, R. K. Mishra and F. S. Emami, *Langmuir*, 2013, **29**, 1754–1765.

S6    J. C. Phillips, R. Braun, W. Wang, J. Gumbart, E. Tajkhorshid, E. Villa, C. Chipot, R. D. Skeel, L. Kalé and K. Schulten, *J. Comput. Chem.*, 2005, **26**, 1781–1802.

S7    D. J. Frisch, M. J.; Trucks, G. W.; Schlegel, H. B.; Scuseria, G. E.; Robb, M. A.; Cheeseman, J. R.; Scalmani, G.; Barone, V.; Petersson, G. A.; Nakatsuji, H.; Li, X.; Caricato, M.; Marenich, A. V.; Bloino, J.; Janesko, B. G.; Gomperts, R.; Mennucci, B.; Hratch, 2016, *Gaussian, Inc., Wallingford CT, 16*.

S8    J. Qi, N. Edirisinghe, M. G. Rabbani and M. P. Anantram, *Phys. Rev. B.*, 2013, **87**, 085404.